\documentclass[lettersize,journal]{IEEEtran}
\usepackage{amsmath,amsfonts}
\usepackage{algorithmicx}
\usepackage{algorithm}
\usepackage{algpseudocode}%
\usepackage{array}
\usepackage[frozencache=true,cachedir=minted-cache]{minted}
\usepackage{enumitem,kantlipsum}
\usepackage{textcomp}
\usepackage{stfloats}
\usepackage{url}
\usepackage{verbatim}
\usepackage{graphicx}
\usepackage{cite}
\usepackage{multirow}
\usepackage[usenames,dvipsnames]{xcolor}
\usepackage[all]{tcolorbox}
\usepackage{tabularx}
\usepackage{colortbl}
\usepackage{caption}
\usepackage{subcaption}
\tcbuselibrary{skins}
\tcbset{tab2/.style={enhanced,fonttitle=\bfseries, fontupper=\scriptsize\sffamily,
colback=blue!5!white,colframe=blue!50!black,colbacktitle=cyan!40!white,
coltitle=black,center title}}

\begin{document}

\title{Embedding-Assisted Attentional Deep Learning for Real-World RF Fingerprinting of Bluetooth}

\author{Anu Jagannath,~\IEEEmembership{Senior Member,~IEEE,} Jithin Jagannath,~\IEEEmembership{Senior Member,~IEEE}
\thanks{Anu Jagannath is with Northeastern University and Marconi-Rosenblatt AI/ML Innovation Laboratory, ANDRO Computational Solutions, LLC.}
\thanks{Jithin Jagannath is with University at Buffalo and Marconi-Rosenblatt AI/ML Innovation Laboratory, ANDRO Computational Solutions, LLC.}
\vspace{- 0.8cm}}


\markboth{Journal of \LaTeX\ Class Files,~Vol.~14, No.~8, August~2021}%
{Shell \MakeLowercase{\textit{et al.}}: A Sample Article Using IEEEtran.cls for IEEE Journals}


\maketitle

\begin{abstract}

A scalable and computationally efficient framework is designed to fingerprint real-world Bluetooth devices. We propose an embedding-assisted attentional framework (Mbed-ATN) suitable for fingerprinting actual Bluetooth devices. Its generalization capability is analyzed in different settings and the effect of sample length and anti-aliasing decimation is demonstrated. The embedding module serves as a dimensionality reduction unit that maps the high dimensional 3D input tensor to a 1D feature vector for further processing by the ATN module. Furthermore, unlike the prior research in this field, we closely evaluate the complexity of the model and test its fingerprinting capability with real-world Bluetooth dataset collected under a different time frame and experimental setting while being trained on another. Our study reveals a $9.17\times$ and $65.2\times$ lesser memory usage at a sample length of 100 kS when compared to the benchmark - GRU and Oracle models respectively. Further, the proposed Mbed-ATN showcases a $16.9\times$ fewer FLOPs and $7.5\times$ lesser trainable parameters when compared to Oracle. Finally, we show that when subject to anti-aliasing decimation and at greater input sample lengths of 1 MS, the proposed Mbed-ATN framework results in a 5.32$\times$ higher TPR, $37.9$\% fewer false alarms, and $6.74\times$ higher accuracy under the challenging real-world setting.
\end{abstract}


\section{Introduction}

\IEEEPARstart{R}{adio} frequency (RF) fingerprint based on the hardware imperfections of the emitter circuit serves as an excellent tool or watermark to distinguish between devices manufactured by the same manufacturer even while transmitting the same message. In the present day and evolving Internet of Things (IoT) era where numerous wireless devices emerge everyday, the wireless security and the privacy of data shared across the spectrum accessed by these devices is a growing concern \cite{device_fingerprint,Ajagannath2022ComST2022}. 

RF fingerprinting pertaining to the extraction of physical layer-level (hardware circuitry) imperfections is emerging as a passive form of wireless emitter identification, i.e., a type of security scheme that can be implemented in a passive wireless receiver without any \textit{apriori} knowledge of the emissions from the emitter. Wireless fingerprinting enables several security-related applications such as indoor positioning \cite{wifi_indoor_1,wifi_indoor_survey,fm_indoor_1}, emitter tracking and localization \cite{ble_tracking_1,ble_tracking_2,BT_tracking_1}, device identification and authentication \cite{dev_identi_1,dev_identi_2}, among others. These applications leverage the signals of opportunity (SoOP) \cite{fm_indoor_1}, i.e., existing RF emissions that hold significant information about their source. For instance, WiFi, Bluetooth (BT), Frequency Modulated (FM) broadcast signals, LoRa, ZigBee, among others are SoOP signals that are ubiquitously present in the spectrum. A passive listener (receiver radio) can therefore decipher and characterize the emitter without sharing any mutual handshake information. However, extracting the emitter features from the overheard signals hold a plethora of unique challenges. 


RF fingerprinting falls under the broader signal intelligence application \cite{Jagannath19MLBook} such as modulation classification \cite{amc_2021,Jagannath18ICC,amc_clement_2021,amc_2022,AJagannath22PHYCOM,amc_res_2021}, signal/protocol classification \cite{radar_rec_2021,JagannathAdHoc2019}, etc. but is considered much more challenging due to the minute nature of the hardware-intrinsic features. A few of those challenges that severely affect RF fingerprinting involve the aging of hardware components, effect of wireless standard (complexity of the waveform) on the fingerprint features, effect of multipath propagation on the received waveform contributed by the obstacles, walls, environmental changes, location changes, noise, interference, etc. For instance, the frequency hopping nature of the BT waveform makes it challenging to capture the hardware-intrinsic features. Here, the performance depends significantly on the input preprocessing, input sample length, DL architecture, among others. This difficult nature was demonstrated in \cite{AJagannath22GLOBECOM} where the single task model performance for the BT fingerprinting was poorer than with WiFi. 
Model generalization challenges were portrayed in \cite{shawabka2020exposing} where even with WiFi channel equalization, their baseline model could achieve only 23.2\% accuracy under the different \textit{day} training and testing (Train One Day and Test on Another - TDTA). These factors consequently affect the performance of the trained and deployed deep learning (DL) models in the real-world operational environment. Hence, studies in this realm \textit{must emphasize validating the generalization capability of the model} such that the effect of these confounding factors is accounted for in the evaluation. To this end, we collect emissions from actual IoT devices in a real-world operational environment to capture the differences in the \textit{time frame, channel, location, and testbed setups} during training and deployment. Hence, the samples are collected under two \textit{different testbed setups with different time frames and locations}.

\section{Related Works}

The various approaches towards RF fingerprinting to enhance the security of wireless devices that utilize wireless standards such as WiFi, BT, and LoRa are an actively researched topic \cite{oracle,lora_rff, bisio2018unauthorizedUAV,AJagannath22GLOBECOM}. The earlier works focused on traditional approaches such as wavelet-based \cite{klein_wavelet}, I/Q imbalance \cite{zhuo2017iqImbalance}, radio turn-on transient-based \cite{yuan2014HilbertHuang} among others \cite{yuan2019mfmcf, deng2017radio, baldini2017PNandDN}. After the recent resurgence of machine learning, DL has been leveraged to overcome some of the challenges \cite{Ajagannath2022ComST2022}. However, the application of DL especially a lightweight deployable framework that improves generalization capability for fingerprinting real-world BT devices is lacking. 

\subsection{Real-World Emitters but Without Generalization Test} 

The authors of \cite{bisio2018unauthorizedUAV} study WiFi-based drone detection with actual drone emitters rather than synthetically generated emissions. In \cite{UAV_multiClass_2020}, a 1D AlexNet and ResNet architectures are adopted to fingerprint 7 DJI M100 drones. The generalization test performed here involves training and testing on different bursts of the emission collected during the same time frame. A ZigBee emitter fingerprinting with Differential Constellation Trace Figure (DCTF) using a LeNet-5 CNN model was proposed in \cite{peng2020DCTF}. The ZigBee emissions are collected, trained, and tested on the same time frame, location, and testbed setup. However, these works do not perform the generalization test where the classifiers are trained with data obtained from \textit{a certain time frame, location, and testbed setup} and tested on another unseen time frame, location, and testbed setup, quoted in our article as TTD scenario. As discussed previously, this will be key to developing reliable real-world RF fingerprinting algorithms. 


\subsection{Studies Considering Generalization Test}

The authors of \cite{shawabka2020exposing} conduct CNN-based WiFi fingerprinting on a custom as well as large-scale DARPA dataset using a channel equalization approach. Here, the authors perform a generalization test where only the time frame of the testing dataset is different from the training, reported in their article as Train on One Test on Another (TDTA) scenario. The authors report an accuracy of only 23.2\% with their Baseline CNN model even with WiFi channel equalization. A LoRa emitter fingerprinting using a spectrogram-based CNN is presented in \cite{lora_rff_shen}. The authors focus on the carrier frequency offset (CFO) of the LoRa emitters in their work. The evaluations to validate model generalization is carried out for different day training and test for a wired setup where the emitters are cabled over an attenuator to the receiving Universal Serial Radio Peripheral (USRP) radio. For the analysis conducted in the wireless setting, only same day training and testing setup is conducted. In \cite{reusmuns2019trust}, the authors employ a triplet loss based CNN model to fingerprint base stations transmitting either of 5G New Radio, LTE, or WiFi waveforms. However, these base stations are software-defined radio (leverages USRP B210) based rather than real-world base stations and emit synthetically generated waveforms with MATLAB's LTE, WLAN, and 5G toolboxes. In this work, the authors perform a generalization test by training and testing on different days. However, the multipath effect, fading, and orientation experienced by the emissions under our challenging TTD scenario is closer to the deployment setting faced in the real-world setting. Note that the proposed Mbed-ATN framework attains a 46.5\% accuracy in fingerprinting the challenging frequency hopping BT emitters under the TTD setting.

\subsection{Spatio-temporal model} 

In \cite{gru_rff_droy}, the authors study the fingerprinting efficacy of recurrent neural networks (RNN) and a hybrid of RNN and CNN on Quadrature Phase Shift Keying (QPSK) waveforms from 8 emitters. However, the study utilized synthetic QPSK waveform generated in software and emitted using USRP radios. In contrast, in order to advance the practical application and demonstrate the usability of fingerprinting algorithm, we adopted 10 COTS IoT emitters that generate commercial standard Bluetooth 5.0 signals. We also demonstrate the implementation of a scalable architecture that benefits from the spatio-temporal feature extraction capability of the attentional module as well as conduct the evaluation in a generalized setting under different propagation conditions. Further, we critically dissect, evaluate, and infer the memory, computational, and fingerprinting performance of the proposed architecture alongside other benchmark models.

While the vast RF fingerprinting literature delves into waveforms such as ADS-B, WiFi, LoRa, and Zigbee \cite{Ajagannath2022ComST2022}, a robust DL based approach to fingerprint BT devices capable of handling unseen configuration is still lacking \cite{AJagannath22GLOBECOM}. The core challenge stems from the rapid frequency hopping nature of the BT. In this work, for the first time, we introduce a unique embedding-assisted attentional framework (Mbed-ATN) for fingerprinting BT emitters and evaluate it in depth. We further demonstrate (with visualization) the challenging nature of the BT waveform in conjunction with realistic deployment conditions under different location, testbed setup, and time frame settings than those the model have been trained with. 

\subsection{Key Contributions}

Unlike existing literature, we comprehensively evaluate the model's complexity, prediction capability, and generalization merit. We measure the generalization power of the proposed DL model by evaluating with unseen data obtained from a different time frame, location, and testbed setup compared to the training data. Our contributions are summarized below,
\begin{itemize}
\item \textbf{Architecture: }We propose for the first time, an embedding-assisted attentional framework for fingerprinting BT devices that provide on average 90.5\% fingerprinting accuracy and demonstrate the lightweight and scalable nature of the proposed DL model to validate its practical deployment capability.
\item \textbf{Dataset:} We collect real-world BT emissions from actual IoT devices under two indoor laboratory scenarios in rich multipath propagation, noise, and scattering (due to obstacles) settings. The datasets utilized in this article have also been published in IEEE dataport \cite{AJagannath2022_BT_port} to foster deployment-friendly research in the fingerprinting realm. Such a challenging dataset that facilitates model generalization validation for BT emitters has never been collected and made available to the public, hence it is considered extremely pivotal in RF fingerprinting research.

\item \textbf{Experimental evaluation: }We present the evaluation results of the proposed DL model in contrast to the benchmark with RF data collected under a different time frame, location, and experimental setup than the training data in addition to analyzing the effects of input tensor length and anti-aliasing filtering.
\item \textbf{Model Analysis:} We further shed light on how to critically evaluate neural network architecture on the basis of their memory consumption and other aspects of model complexity that are essential to building models for practical deployment. The proposed model consumes $65.2\times$ lesser memory, $16.9\times$ fewer FLOPs, and $7.5\times$ fewer trainable parameters in contrast to the state-of-the-art architecture. 
\end{itemize}

\section{embedding-assisted RF Fingerprint Extractor}
\label{sec:mbed_atn}
\begin{figure*}[htp]
\centering
\includegraphics[width=1.98\columnwidth]{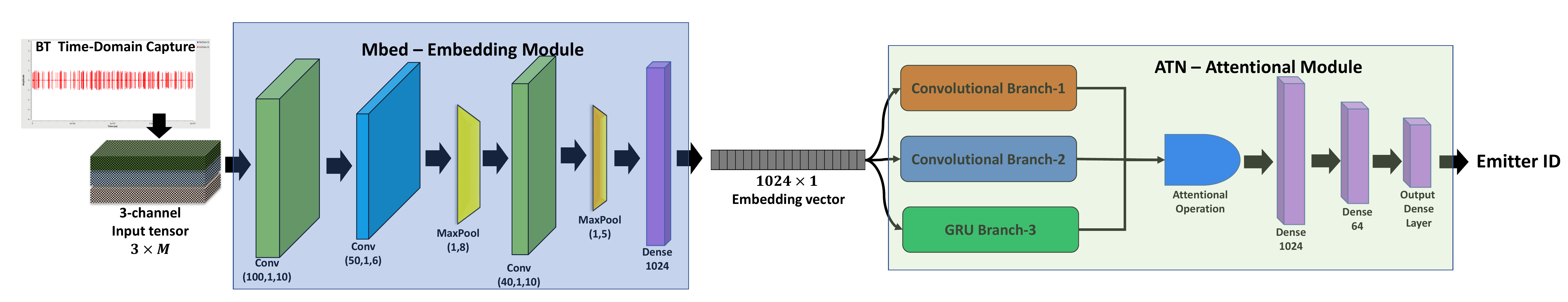} 
\caption{Proposed Scalable Mbed-ATN framework}
\label{fig:mbed-atn}
\end{figure*}

In this section, we elaborate on the design of the proposed embedding-assisted RF fingerprint extractor (Mbed-ATN) enabling it to classify BT emitters in an unseen challenging environment. The proposed Mbed-ATN is a deep learning framework that adopts a convolutional neural network (CNN)-based embedding module (Mbed) that serves as the feature extractor and dimensionality reduction module. The Mbed module maps the high dimensional BT signal input tensor ($3\times M, M =$ [10k, 100k, 1M]) to a one dimensional (1D) $1024\times1$ vector which feeds into a CNN and gated recurrent unit (GRU)-based attentional (ATN) classifier. The ATN module extracts the spatial and sequential patterns in the input vector allowing it to efficiently isolate the fingerprint from other confounding factors. This unique Mbed-ATN framework that combines the advantages of CNN and GRU in extracting the unique emitter characteristics is shown in Fig. \ref{fig:mbed-atn}. Emphasizing the significance of deployment of the Mbed-ATN in real-world operational scenarios, we enforce a lightweight and scalable architecture that can generalize well to the real-world environment. 

\textbf{Input Data Preprocessing}: We denote the time domain BT signal of length $N$ samples captured by the receiver as $y(t)|_{i=1}^N$. In our previous work \cite{AJagannath22GLOBECOM} and as an ongoing study, we have empirically determined that BT emitter fingerprinting requires larger input sample lengths and additional features in the input tensor for acceptable classification accuracy. The capture length in this study is intentionally kept large enough ($N = 40$ MS) to experiment with data segmentation and other signal processing required to determine the input format that yields an acceptable fingerprinting accuracy. 
We subject the captured BT signal to the following operations to generate a $3\times M$ tensor.
\begin{align}
\scriptsize
    \mathbf{Y}^{3\times M} &=  \;  \mathfrak{F}\big[\hat{y}(t)_{t=1}^M\big] \\
    &= \; \begin{bmatrix} |\hat{y}(t)_{t=1}^M|\\
    \angle \hat{y}(t)_{i=1}^M\\
    PSD\bigg(\hat{y}(t)_{t=1}^M\bigg)\end{bmatrix}\label{eq:ip_format}
\end{align}
where $\hat{y}(t)_{t=1}^M$ is the downsampled version of $y(t)_{t=1}^N$, the first two rows contain the magnitude and phase of the decimated signal $\hat{y}(t)$ and the third row is the power spectral density (PSD) of the decimated signal.

\textbf{Embedding Module:} We resort to the powerful feature extraction capability of CNNs to process the input tensor $\mathbf{Y}$. The Mbed module acts as a dimensionality reduction step in mapping the large 3D input tensor to a condensed 1D feature-embedded vector. It treats the 3D input tensor as a 3-channel input and adopts 1D convolutional kernels to encode the dependencies between the adjacent samples in each input channel. 
The architectural detail of the Mbed module is shown in Table \ref{tab:mbed}. We resort to using the parametric ReLU (PReLU) activation function \cite{prelu} in the convolutional layers as it has shown considerable improvement when the negative values are not zeroed out. The PReLU performs non-linear mapping of an input $x$ as in equation \ref{eq:prelu}.
\begin{align}
\small
    f(x) = \begin{cases} x, & \text{if } x>0\\
            ax, & \text{if } x\leq 0
    \end{cases}\label{eq:prelu}
\end{align}
Here, $a$ is the trainable parameter and hence, the name PReLU. The dense layer utilizes ReLU activation. Unlike PReLU, the ReLU maps all negative values to 0, or in other words, when the $a=0$ in equation \ref{eq:prelu}, the function is equivalent to ReLU.

\begin{table}[h!]
\renewcommand{\arraystretch}{1.2}
\begin{center}
\scriptsize
\caption{Architectural detail of Mbed module. \label{tab:mbed}}
\begin{tabular}{|p{7.5 cm}|} 
 \hline
 Input $3\times M$ \\ \hline
 Conv (100,1,10) - Stride (1,10) $\forall M<1e6$ - Stride (1,20) $\forall M\geq1e6$\\ \hline
 Conv (50,1,6) - Stride (1,3) $\forall M<1e6$ - Stride (1,6) $\forall M\geq1e6$  \\ \hline
 Maxpool (1,8) Dropout 0.5\\\hline
 Conv (40,1,10) - Stride (1,10) $\forall M<1e6$ - Stride (1,5) $\forall M\geq1e6$  \\ \hline
 Maxpool (1,5) - active for $M\geq1e6$ Dropout 0.5\\ \hline
 Dense 1024 \\\hline
 \textbf{Activation:} Conv Layers - PReLU, Dense Layer - ReLU\\\hline
\end{tabular}
\end{center}
\end{table}

\textbf{Attentional module:} We resort to adopting an attentional mechanism to extract the inter-dependencies in the samples and \textit{focus only on the relevant portions of the samples} with fewer layers. Our past experience and experiments show that the adoption of the attentional module can outperform deep network architectures and preempt the need for denser networks. The ATN module is a hybrid model that combines the benefits of CNN and GRU. While the CNN performs 1D convolutions on the embedding vector $\mathbf{f}$ to capture the timing relationship of the samples, the GRU extracts the before-after timing dependencies of the samples. We consider this as a pivotal step in characterizing and comprehensively extracting the fingerprint features, especially owing to the hopping nature of the BT waveform as it traverses the multipath propagation channel.
\begin{table}[h!]
\renewcommand{\arraystretch}{1.2}
\begin{center}
\caption{Architectural detail of ATN module. \label{tab:atn}}
\begin{tabular}{|p{8.3 cm}|} 
 \hline
 Input $1024\times 1$ \\ \hline
 \textbf{Branch-1:} Conv (15,1,7) - Stride 1 - Padding 1 - PReLU - Dropout 0.1 \newline 
 Conv (32,1,7) - Stride 1 - PReLU - MaxPool (1,2) - Dropout 0.5 \\ \hline
 \textbf{Branch-2:} Conv (15,1,3) - Stride 1 - Padding 1 - PReLU - Dropout 0.1 \newline 
 Conv (32,1,3) - Stride 1 - PReLU - MaxPool (1,2) - Dropout 0.5 \\ \hline
\textbf{Branch-3:} GRU hidden size = 80, \#layers = 3, Dropout = 0.5 \\
\hline
Dense 1024 - PReLU - Dropout 0.2 \\\hline
Dense 64 - PReLU - Dropout 0.2 \\\hline
Dense 10 - Softmax\\\hline
\end{tabular}
\end{center}
\end{table}
GRU is an efficient form of long short term memory (LSTM) since it uses only two gates - \textit{Update} and \textit{Reset} - instead of three gates as in LSTM. Further, GRU does not possess an internal memory or an output gate. Therefore, GRU uses fewer training parameters and memory and hence trains faster than LSTM. The update ($\textbf{u}_t$) gate controls the amount of past information that needs to be carried over to the next state. The reset ($\textbf{r}_t$) gate determines the amount of previous history that needs to be forgotten. The GRU units are defined by the following set of equations,
\begin{align}
\small
    \textbf{u}_t &= \sigma\big(\textbf{W}_u\textbf{x}_t + \textbf{R}_u\textbf{h}_{t-1}+\textbf{b}_u \big)\\
    \textbf{r}_t &= \sigma\big(\textbf{W}_r\textbf{x}_t + \textbf{R}_r\textbf{h}_{t-1}+\textbf{b}_r  \big)\\
    \hat{\textbf{h}}_t &= \tanh\big(\textbf{W}_h\textbf{x}_t + \textbf{R}_h\big(\textbf{r}_t\odot \textbf{h}_{t-1}\big)+\textbf{b}_h  \big)\\
    \textbf{h}_t &= (1-\textbf{u}_t)\odot \textbf{h}_{t-1} + \textbf{u}_t\odot \hat{\textbf{h}}_t
\end{align}
where $\textbf{x}_t$ is the input vector, $\textbf{W}_i$ and $\textbf{R}_i$ are the weight matrices, $\textbf{b}_i$ the bias vector, $\textbf{h}_t$ indicates candidate hidden state, $\tanh(\circ)$ is the hyperbolic tangential activation function, and $\sigma(\circ)$ is the sigmoid activation function.

As in the architectural diagram in Fig.\ref{fig:mbed-atn}, the input ($\textbf{f}$) to the ATN module feeds into two convolutional branches, and a GRU branch. The notations $\mathcal{C}_1$, $\mathcal{C}_2$, and $\mathcal{G}$ denote the operations of the first convolutional branch, second convolutional branch, and the GRU branch respectively. The layer details of the branches are presented in Table \ref{tab:atn}. The output from the convolutional branches is vectorized (flattened) form of their respective feature maps. The GRU branch is a many-to-1 type of GRU whose output is also a vector. The operations of the ATN module are governed by the following set of equations,

\begin{align}
\small
    \textbf{o}_1 &= \mathcal{C}_1\big(\textbf{f}\big) &\textbf{s}= \text{SiLU}\big(\textbf{o}_3\big)\\
    \textbf{o}_2 &= \mathcal{C}_2\big(\textbf{f}\big)  &\textbf{a} = \big[\textbf{o}_1;\textbf{o}_2;\textbf{s} \big]\label{eq:a} \\
    \textbf{o}_3 &= \mathcal{G}\big(\textbf{f}\big)
\end{align}

where $\textbf{s}$ is the scoring vector function approximation obtained by applying Sigmoid Linear Unit (SiLU) activation to the output from the GRU branch. The SiLU activation \cite{silu} multiplies the input ($x$) by its sigmoid activation ($\sigma(x)$). The operator $;$ indicates vector concatenation. The final attentional vector $\textbf{a}$ is generated by concatenating the outputs from the convolutional branches with the scoring vector $\textbf{s}$ as in equation (\ref{eq:a}). This scoring vector is fed into the subsequent Dense layers for the final softmax emitter classification.

\textbf{Training Mbed-ATN framework:} We train the end-to-end Mbed-ATN framework as in Algorithm 1. The Mbed module is initially trained to classify the emitters by adding an output softmax Dense layer to the architecture in Table \ref{tab:mbed}. This layer is dropped after training and the $1024\times1$ feature vector $\textbf{f}$ is fed to the ATN module. The ATN module is then trained independently while keeping the weights of the Mbed module unchanged. The modules are trained for maximum epochs of 2000 with Adam optimizer at a learning rate of 0.0001. The network convergence is monitored during the training process and the parameters are frozen at the best point of convergence.

\begin{algorithm}
\small
\caption{Backpropagation to train Mbed-ATN framework}\label{alg:train}
\begin{algorithmic}
\State \textbf{Train Mbed module}:
\State Initialize network weights $\Theta_{Mbed}$.
\For{\texttt{epoch = 1 to MAX\_EPOCHS}} 
\For{\texttt{steps = 1 to STEPS}}
\State \textbf{Input} batch $\mathbf{x}$ and \textbf{Compute} loss \State $\ell_{Mbed}(\Theta_{Mbed})$ [standard forward pass]
\State \textbf{Compute} gradients $\nabla \ell_{Mbed}(\Theta_{Mbed})$
\State \textbf{Update} weights \State $\Theta_{Mbed}^{*}\longleftarrow\Theta_{Mbed}$ [standard backward pass]
\EndFor
\State Stop training once model stops learning (starts to diverge)
\EndFor
\State Freeze the Mbed module with learned weights $\Theta_{Mbed}^{*}$
\State Eliminate the output softmax Dense layer of Mbed module and \State feed the $1024\times1$ feature vector $\textbf{f}$ to ATN module.
\State \textbf{Train Mbed-ATN module}:
\State Initialize network weights $\Theta_{Mbed}^{*}, \Theta_{ATN}$.
\For{\texttt{epoch = 1 to MAX\_EPOCHS}} 
\For{\texttt{steps = 1 to STEPS}}
\State \textbf{Input} batch $\mathbf{x}$ and \textbf{Compute} loss \State $\ell_{Mbed-ATN}(\Theta_{Mbed}^{*}, \Theta_{ATN})$ [standard forward pass]
\State \textbf{Compute} gradients $\nabla \ell_{Mbed-ATN}(\Theta_{Mbed}^{*}, \Theta_{ATN})$
\State \textbf{Update} weights of ATN module\State $\Theta_{ATN}^{*}\longleftarrow\Theta_{ATN}$ [standard backward pass]
\EndFor
\State Stop training once model stops learning (starts to diverge)
\EndFor
\end{algorithmic}
\end{algorithm}


\section{Experimental Evaluation}
\textbf{Real-world IoT Datasets:} We consider a testbed with real-world commercial IoT devices for the practical application and evaluation of the proposed RF fingerprinting framework. We collect the BT emissions from 10 IoT emitters in two challenging settings in an indoor multipath environment with other unavoidable interferences and obstacles rendering a rich multipath propagation scenario. A passive listener USRP X300 tuned into a 2 MHz bandwidth of a 2.414 GHz center frequency is streaming samples at the rate of 2 MS/s. The USRP X300 is outfitted with a UBX160 daughterboard and a VERT2450 antenna affixed to the RX2 antenna port. The respective emitters are positioned to transmit the BT bursts throughout the capture while the receiving radio records $40$ MS in one capture. For the benefit of research in the domain, the datasets utilized in this article have also been published in IEEE dataport \cite{AJagannath2022_BT_port}.

From the RF fingerprinting-specific feature extraction perspective, the BT waveform is challenging in itself owing to the frequency hopping nature which hops at the rate of 1600 hops/second over the 2.4 GHz ISM band. This implies the signal will be periodically visiting the tuned in BT channel making it a harder waveform to capture and fingerprint. The shorter input sample lengths therefore cannot comprehensively capture the emitter characteristics and will therefore need larger sample lengths \cite{AJagannath22GLOBECOM}.

\textbf{\textit{Setup 1:}} Here the emitters and the receiving radio are positioned in line-of-sight (LoS) settings. The separation between the emitter and receiver is varied from $1.6$ ft to $9.8$ ft in steps of $0.8$ ft.

\begin{figure}[htp]
\centering
\includegraphics[width=.99\columnwidth]{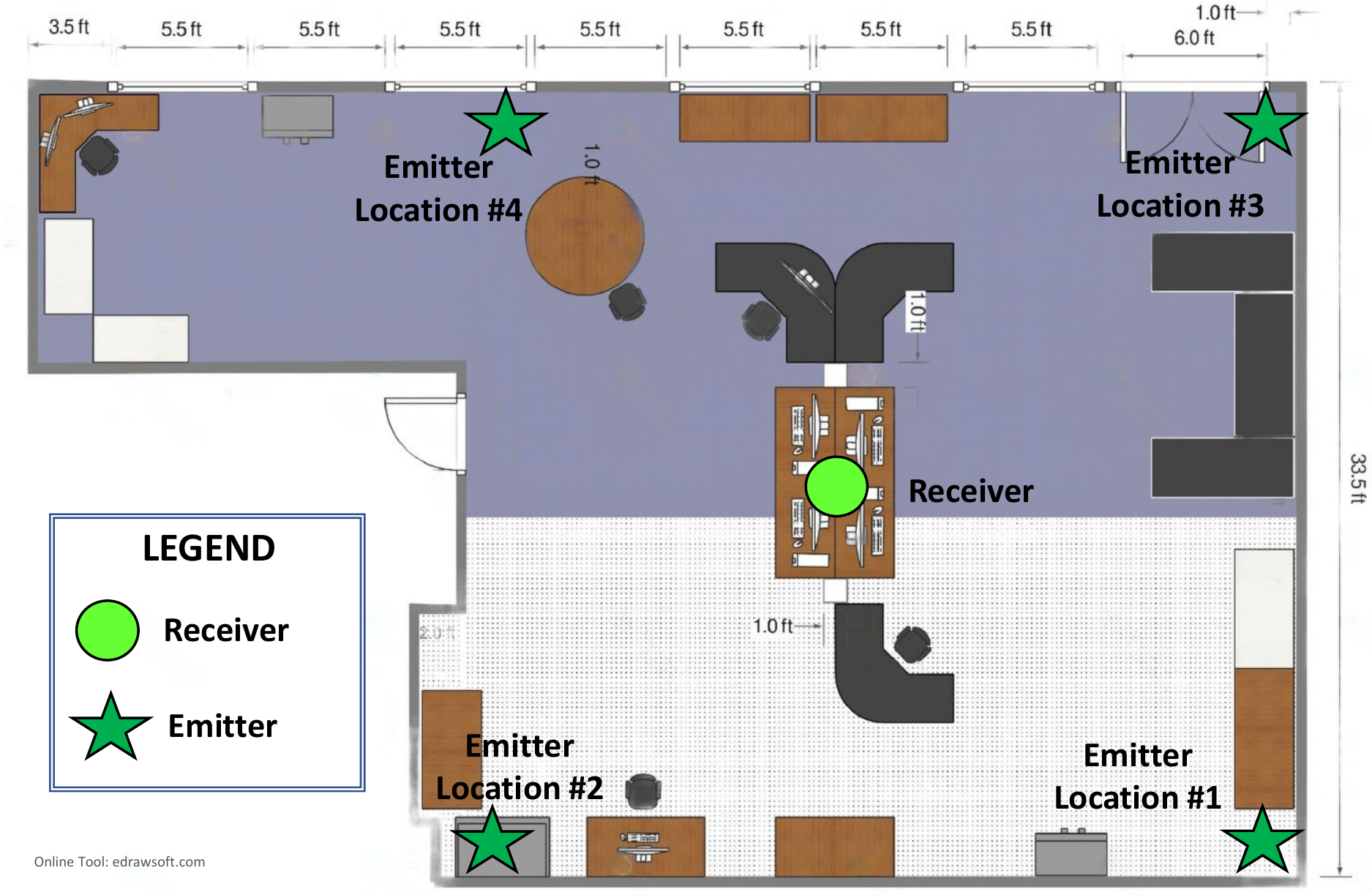} 
\caption{Testbed layout for Setup 2 demonstrating the emitter in one of the corners and receiver at the center of the laboratory.}
\label{fig:corner}
\end{figure}

\textbf{\textit{Setup 2:}} This setup is considered a challenging setting given the rich multipath propagation settings between the emitters and the receiver. Here the emitter is placed at the four corners of the indoor laboratory while the receiver is placed at the center of the laboratory space. In this setup, the maximum separation between one of the corners and the receiver amounts to approximately $24.2$ ft. Figure \ref{fig:corner} shows the indoor laboratory setup with emitter locations and receiver placement. The multipath nature of the laboratory contributed by the walls, furniture, and lab equipment are clearly portrayed here (to keep the figure less crowded some obstacles are not shown).

\begin{figure}[h!]
     \centering
     \begin{subfigure}[b]{0.23\textwidth}
         \centering
         \includegraphics[width=1 \textwidth]{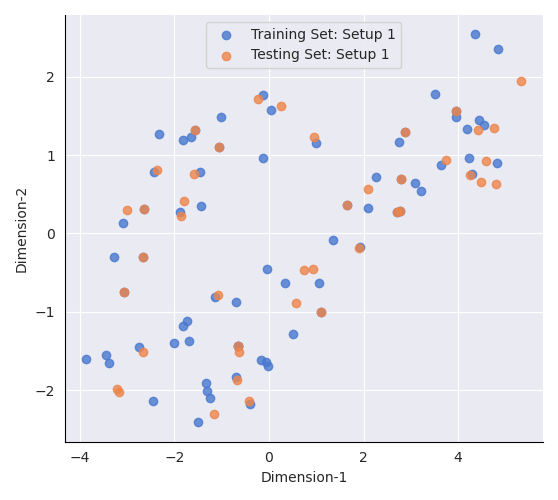}
         \caption{TTS Distribution}
         \label{fig:ttsd_tsne}
     \end{subfigure}
     \hfill
     \begin{subfigure}[b]{0.23\textwidth}
         \centering
         \includegraphics[width=1 \textwidth]{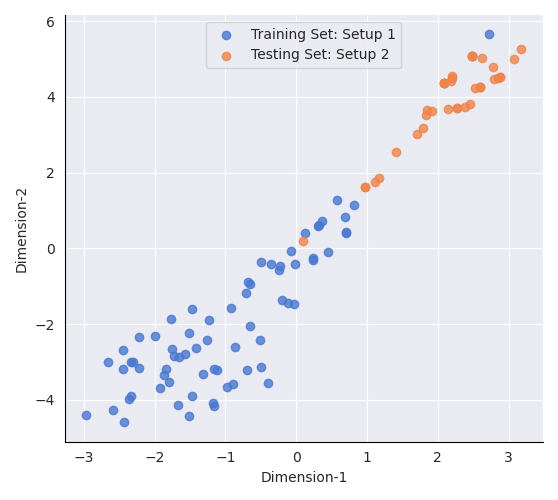}
         \caption{TTD Distribution}
         \label{fig:ttddl_tsne}
     \end{subfigure}
     \hfill
        \caption{t-SNE visualization of training and testing data distribution under the TTS and TTD evaluation settings.}
        \label{fig:tsne}
\end{figure}

\textit{\textbf{Generalization Evaluation:}} We consider two categories of experimental evaluations to quantify the generalization capability of the model to suit real-world deployment.
\begin{enumerate}
    \item Train Test Same time frame, location, and testbed setup (TTS) - Here, the samples for the training and testing set are drawn from the data captured in the same time frame, location, and testbed setup. For the TTS scenario, we use the samples from the Setup 1 for training and testing.
    
    \item Train Test Different time frame, location, and testbed setup (TTD) - In this scenario, the model is trained with samples collected from Setup 1 and later tested on captures from Setup 2. 
\end{enumerate}
We validate the argument that the training and testing data under the TTD have different distributions by utilizing the well-known t-Distributed Stochastic  Neighbor Embedding (t-SNE) visualization tool. The training and testing data distribution under the TTS and TTD setting are shown in Fig.\ref{fig:ttsd_tsne} and Fig.\ref{fig:ttddl_tsne} respectively. Figure \ref{fig:ttsd_tsne} indicates same distributions for TTS unlike TTD (Fig.\ref{fig:ttddl_tsne}) where the samples possess different features/distributions.

\subsection{Key Performance Indicators}

In this section, we specify the performance metrics that are used to evaluate the models. The predictions made by any DL model or in other words, the confusion matrix can be categorized into true positives (TP), true negatives (TN), false positives (FP), and false negatives (FN).
\begin{enumerate}[leftmargin=*]
    \item \textit{True positive rate (TPR) or Recall}: quantifies the positive predictions made by the model with respect to total positive predictions. For a multi-class classification, it is $TPR = \frac{\sum_i TP_i}{\sum_i (TP_i + FN_i)}$, where $i$ denotes the class $i$.
    \item \textit{False positive rate (FPR)}: measures the false predictions of the model in proportion to the total false predictions. Its computed as $FPR = \sum_{i=1}^{L} \frac{FP_i/(FP_i + TN_i)}{L}$.
    \item T\textit{op-1 accuracy} (or balanced accuracy): is the arithmetic mean of the recall for each class.
    \item \textit{FLOPs}: accounts for the total number of floating point operations in the model.
    \item \textit{Model parameters}: measures the total number of trainable parameters in the model.
    \item \textit{Supported sample lengths}: the maximum measured input tensor lengths supported by the model without causing any out-of-memory (OOM) GPU errors.
\end{enumerate}

\subsection{Complexity Analysis}

Model complexity is an often overlooked factor by developers while designing and training deep learning (DL) models. According to a recent empirical study on 4960 failed DL jobs in Microsoft, 8.8\% of the job failures were caused due to the depletion of GPU memory accounting for the largest category in all DL specific failures \cite{microsoft}. The frequency hopping nature of the BT waveform requires a scalable architecture that can process larger input sample lengths. This makes the model architecture challenging since it must be \textit{large enough to process larger input samples but at the same time be lightweight for supporting commercial-off-the-shelf (COTS) deployment platforms}. In this section, we perform a systematic review of the memory footprint of the proposed Mbed-ATN model and benchmark it against Oracle \cite{oracle} and the GRU-based network proposed in \cite{gru_rff_droy}.

\begin{figure}[h!]
     \centering
     \begin{subfigure}[b]{0.4\textwidth}
         \centering
         \includegraphics[width=\textwidth, height=5 cm]{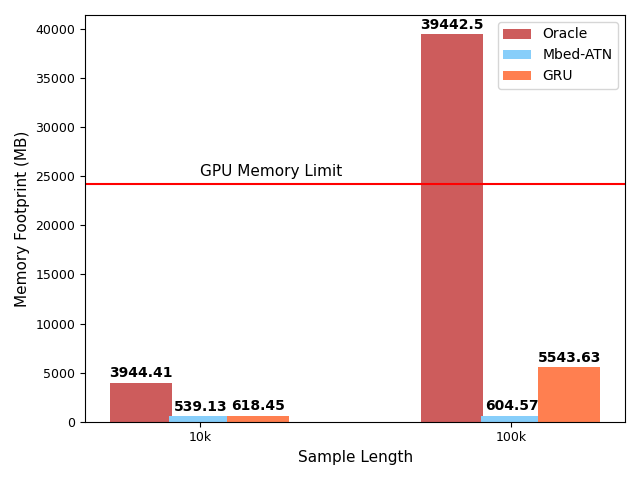}
         \caption{GPU Memory Consumption of training Oracle and Proposed Mbed-ATN models with different sample lengths and Batch size=2.}
         \label{fig:mem_oracle_atn}
     \end{subfigure}
     \hfill
     \begin{subfigure}[b]{0.4\textwidth}
         \centering
         \includegraphics[width=\textwidth, height=5 cm]{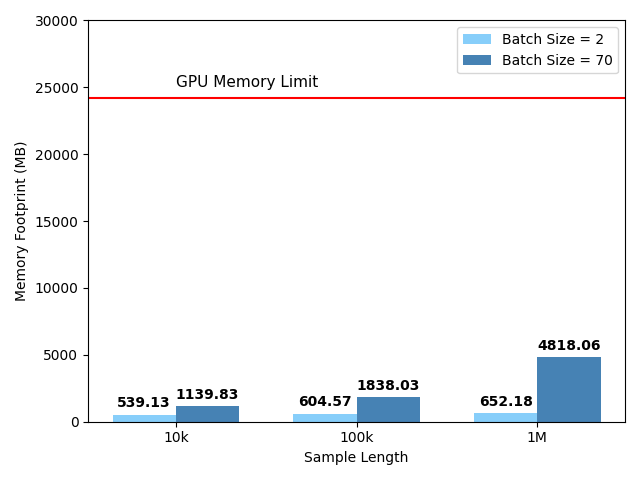}
         \caption{GPU Memory Consumption of training the proposed Mbed-ATN and benchmark models with different sample lengths and batch sizes.}
         \label{fig:mem_atn}
     \end{subfigure}
     \hfill
        \caption{GPU Memory Consumption of training the proposed Mbed-ATN and benchmark models. The red line indicates the memory capacity of the Quadro RTX 6000 GPU.}
        \label{fig:mem}
\end{figure}

Having a firm grasp of the memory usage of a model is imperative in designing efficient and lightweight DL models. In order to answer this critical practical usage question, we elucidate the maximum memory consumption of a model and demonstrate it on the proposed Mbed-ATN and benchmark models. The training of a DL model can be segmented into roughly five stages:
\begin{enumerate}[leftmargin=*]
    \item \textbf{Model Loading:} This stage involves moving the model parameters to the GPU memory. Here the current memory usage is the model memory.
    \item \textbf{Forward pass:} Here the input batch is passed through the model and the intermediate activations are stored in memory for use by backpropagation. Here the current memory consumption is contributed by the model and the activations.
    \item \textbf{Backward pass:} The gradients are computed from the end of the network to the beginning while discarding the saved activations during the traversal. The memory usage in this step is by the model and the gradients.
    \item \textbf{Optimizer parameters:} The optimizer parameters are updated during the backpropagation. The parameters would vary depending on the type of learning algorithm such as Adam, RMSProp, etc. For example, Adam would estimate the first and second moments of the gradients. Here the memory is depleted by the model, gradients, and gradient moments.
    \item \textbf{Training iterations:} Once the first iteration has passed and the optimizer has taken a step, the gradient and gradient moments are updated and stored in memory. So the maximum memory consumption in the subsequent training iterations will be in parts by the model, activations, gradients, and gradient moments.
\end{enumerate}
We used the PyTorch framework and a Quadro RTX 6000 GPU in implementing and evaluating the models. Figure \ref{fig:mem} demonstrates the GPU memory usage by Mbed-ATN model and benchmark models under the same evaluation settings on the challenging frequency hopping BT emissions. In Fig.\ref{fig:mem_oracle_atn}, we analyze the GPU memory usage when the input tensor lengths are configured to $M=10$ kS and $M=100$ kS. It can be seen that the memory usage rapidly scales up to trigger OOM with the Oracle model while the proposed Mbed-ATN maintains manageable and very low memory usage. While the GRU network has a lesser memory usage than Oracle it is still higher and computationally much slower than the proposed Mbed-ATN rendering it infeasible for practical BT fingerprinting applications. To give the reader a quantifiable comparison, at the sample length of $M=100$ kS, the GRU network has an inference time of $\sim6.7$s. Such slow computation is inherent to RNN which processes samples within each example sequentially (output at each step depends on the previous) unlike CNN that can perform parallel computation. This corresponds to a $7.3\times$ and $65.2\times$ lesser memory usage with Mbed-ATN architecture in contrast to Oracle at sample lengths of $M=10$ kS and $M=100$ kS respectively. Similarly, Mbed-ATN scores a $1.15\times$ and $9.17\times$ lesser memory usage at sample lengths of $M=10$ kS and $M=100$ kS respectively when compared to GRU network. This evaluation was carried out with a batch size of 2. A higher batch size of 70 and an input sample length $M=1$ MS were not feasible with the benchmark models. However, to provide more insight to the readers, we characterize the proposed Mbed-ATN at different batch sizes and sample lengths in Fig.\ref{fig:mem_atn}. These analyses demonstrate the GPU memory usage with the proposed Mbed-ATN well under the GPU memory capacity.

To shed more light on the model complexity from a deployment standpoint, we also evaluate the floating point operations (FLOPs), the number of trainable parameters, and inference time of the proposed Mbed-ATN and benchmark models for an input tensor length of $M=10$ kS (Table \ref{tab:flops}). The proposed Mbed-ATN showcases a $16.9\times$ fewer FLOPs and $7.5\times$ lesser trainable parameters when compared to Oracle. Although the GRU network possesses fewer trainable parameters and FLOPs in contrast to both Mbed-ATN and Oracle, the fully recurrent nature of the network renders it computationally much slower as showcased by an inference time which is $21.21\times$ and $31.3\times$ that of the Mbed-ATN and Oracle models respectively. We clarify here that the slightly higher inference time of the proposed Mbed-ATN in contrast to Oracle despite the significantly lesser computational and memory requirement is attributed to the GRU unit employed in the ATN module of the architecture that processes samples present in each input sequentially. Recall here that the FLOPs and inference time are evaluated under the same settings for all models and at a sample length of 10 kS. These experiments demonstrate the superior lightweight nature of the proposed Mbed-ATN model in terms of memory footprint and computational requirements while being capable of supporting a larger input sample length of $1$ MS. Further, it also showcases the scalability limitation of the benchmark models.
\begin{table}
\centering
\caption{FLOPs analysis with the benchmark. \label{tab:flops}}
\begin{tcolorbox}[width=8.6 cm,tab2,tabularx={p{1.3 cm}|p{0.8 cm}|p{1.5 cm}|p{1.7 cm}|p{2 cm}}]
\textbf{Model} & \textbf{FLOPs} & \textbf{\#Parameters} & \textbf{Supported \newline Sample length}&\textbf{Inference \newline Time}\\
\hline
Mbed-ATN &2.181G &33.951M &1M &1.4ms\\
Oracle\cite{oracle} &36.87G &256M &10k &0.95ms\\
GRU\cite{gru_rff_droy} &0.01G &3.627M &10k &29.7ms\\\hline
\end{tcolorbox}
\vspace{-.2 cm}
\end{table}
\subsection{Effect of Sample length}
In this section, we critically evaluate how the length of the input tensor affects the performance of the fingerprinting framework. As mentioned previously, the capture length is $N=40$ MS which is subsequently decimated to different sample sizes ($M$) such as $10$ kS and $1$ MS. In these evaluations, we also characterize the Mbed module's fingerprinting performance separately to showcase the need for the ATN unit.


Since it is only feasible to support (for the given hardware) a sample length of $M = 10$ kS with the benchmark models, we also present its fingerprinting performance in Table \ref{tab:BT10k}. We measure the performance of the models when they are trained and validated with the dataset collected using \textit{Setup 1} and tested with a portion of the test set obtained from unseen data collected from the same scenario, i.e., TTS. Under the TTS evaluation, the proposed Mbed model outperforms the Mbed-ATN as well as benchmark models in terms of the TPR, FPR, and Top-1 accuracy. The performance of Mbed and Mbed-ATN was measured for a sample length of 1 MS Table \ref{tab:BT1M}. Comparing the KPIs of these evaluations under the TTS scenario, demonstrate an increase in TPR and Top-1 accuracy while lower FPR at higher sample lengths. These evaluations further portray that the significance of the ATN module comes into play at a larger sample length ($M=1$ MS). This is intuitive as the GRU branch of the proposed Mbed-ATN can decipher the time series relation better with longer sequence lengths.  Table \ref{tab:BT1M} shows a $2.8$\% higher TPR, $23$\% fewer false alarms, and $0.5$\% greater top-1 accuracy with Mbed-ATN in contrast to the Mbed unit alone under the TTS condition.

\begin{table}
\centering
\caption{Fingerprinting performance at $M = 10$ kS. \label{tab:BT10k}}
\vspace{-.3 cm}
\begin{tcolorbox}[width=8.1 cm,tab2,tabularx={p{1.1 cm}|p{1.6 cm}|p{0.8 cm}|p{0.8 cm}|p{0.8 cm}}]

\textbf{Scenario} &\textbf{Model} & \textbf{TPR} &\textbf{FPR} & \multicolumn{1}{c}{\textbf{Top-1 Acc.}} \\
\hline
\multirow{3}*{TTS}
                    & Mbed &\textbf{0.762} &\textbf{0.027} &\textbf{0.775}\\
                    &Mbed-ATN &0.738 &0.029 &0.742\\
                    &Oracle\cite{oracle} &0.738 &0.029 &0.742\\
                    &GRU \cite{gru_rff_droy}    &0.19 &0.089 &0.211\\\hline
\multirow{3}*{TTD}&Mbed &\textbf{0.158} &\textbf{0.094} &\textbf{0.145}\\
                    &Mbed-ATN &0.079 &0.103 &0.075\\
                    &Oracle\cite{oracle} &0.079 &0.103 &0.069\\
                    &GRU \cite{gru_rff_droy}    &0.157 &0.095 &0.135\\\hline
\end{tcolorbox}
\end{table}

\begin{table}
\centering
\caption{Fingerprinting performance at $M = 1$ MS. \label{tab:BT1M}}
\begin{tcolorbox}[width=8.1 cm,tab2,tabularx={p{1.1 cm}|p{1.6 cm}|p{0.8 cm}|p{0.8 cm}|p{0.8 cm}}]
\textbf{Scenario} &\textbf{Model} & \textbf{TPR} &\textbf{FPR} & \multicolumn{1}{c}{\textbf{Top-1 Acc.}} \\
\hline
\multirow{3}*{TTS}&Mbed &0.881 &0.013 &0.885\\
                    &Mbed-ATN &\textbf{0.905} &\textbf{0.01} &\textbf{0.91}\\\hline
\multirow{3}*{TTD}&Mbed &0.105 &0.095 &0.175\\
                    &Mbed-ATN &\textbf{0.211} &\textbf{0.086} &\textbf{0.275}\\\hline
\end{tcolorbox}
\vspace{-.2 cm}
\end{table}
\subsection{Generalization on different location, testbed, and time frame setting.}
An important aspect of training and deploying a DL model for real-world applications is its generalization capability. The fingerprinting literature has often resorted to evaluating this in terms of train one day and test another (TDTA) scenario where the DL model is trained and validated with one dataset while evaluating it with a test set collected on a different day (time frame) \cite{shawabka2020exposing}. However, unlike the past works, we make it even more challenging by \textit{testing on data collected from not just a different time frame but also under a different location and testbed setup}, i.e., TTD. Under the TTD setting, the models are trained and validated with data collected from \textit{Setup 1} and tested on data captured from \textit{Setup 2}.

The TTD evaluation in Table \ref{tab:BT10k} shows marginally higher performance of the proposed Mbed unit in contrast to the benchmark models. 
The significance of the ATN module is depicted in Table \ref{tab:BT1M} where it achieves $2.6$\% higher top-1 accuracy in contrast to the Mbed unit.

\subsection{Effect of applying anti-aliasing decimation}

\begin{table}
\centering
\caption{Fingerprinting performance with anti-aliasing decimation (sample length = 1M). \label{tab:BTanti}}
\begin{tcolorbox}[width=8.1 cm,tab2,tabularx={p{1.1 cm}|p{1.6 cm}|p{0.8 cm}|p{0.8 cm}|p{0.8 cm}}]
\textbf{Scenario} &\textbf{Model} & \textbf{TPR} &\textbf{FPR} & \multicolumn{1}{c}{\textbf{Top-1 Acc.}} \\
\hline
\multirow{3}*{TTS}&Mbed &0.905 &0.011 &0.905\\
                    &Mbed-ATN &0.905 &0.011 &0.905\\\hline
\multirow{3}*{TTD}&Mbed &0.342 &0.072 &0.395\\
                    &Mbed-ATN &\textbf{0.421} &\textbf{0.064} &\textbf{0.465}\\\hline
\end{tcolorbox}
\vspace{-.2 cm}
\end{table}

\begin{figure*}[h!]
     \centering
          \begin{subfigure}[b]{0.32\textwidth}
         \centering
         \includegraphics[width=\textwidth, height=5 cm]{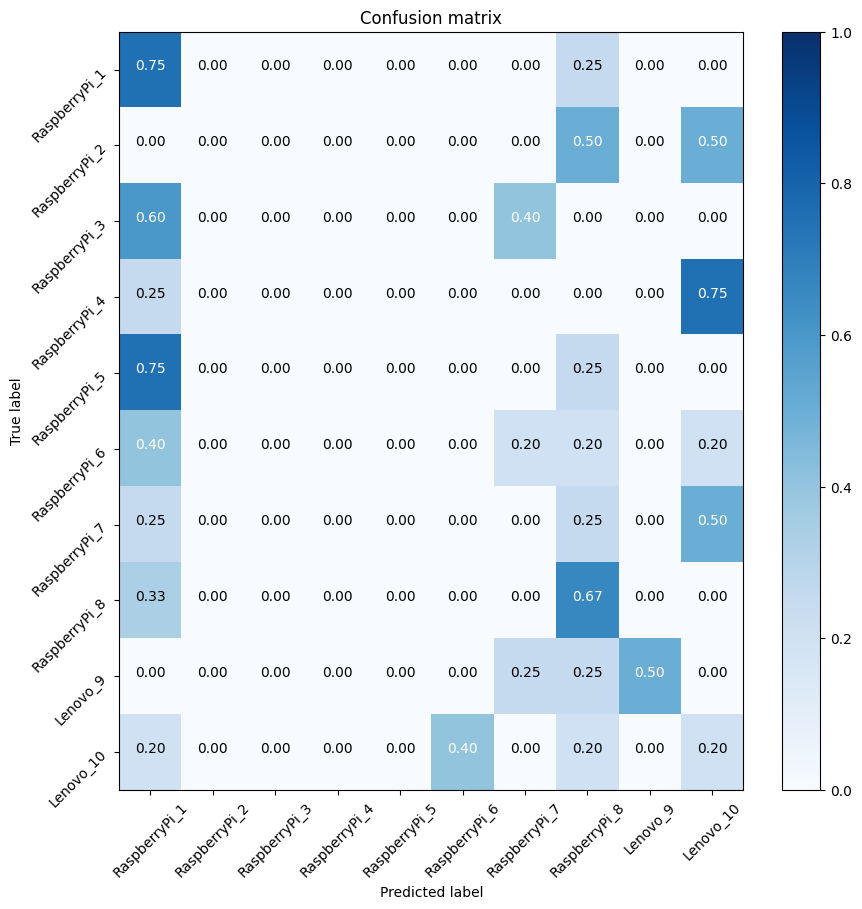}
         \caption{GRU network \cite{gru_rff_droy} performance with the maximum supported sample length}
         \label{fig:gru_ttsd}
     \end{subfigure}
     \hfill
     \begin{subfigure}[b]{0.32\textwidth}
         \centering
         \includegraphics[width=\textwidth, height=5 cm]{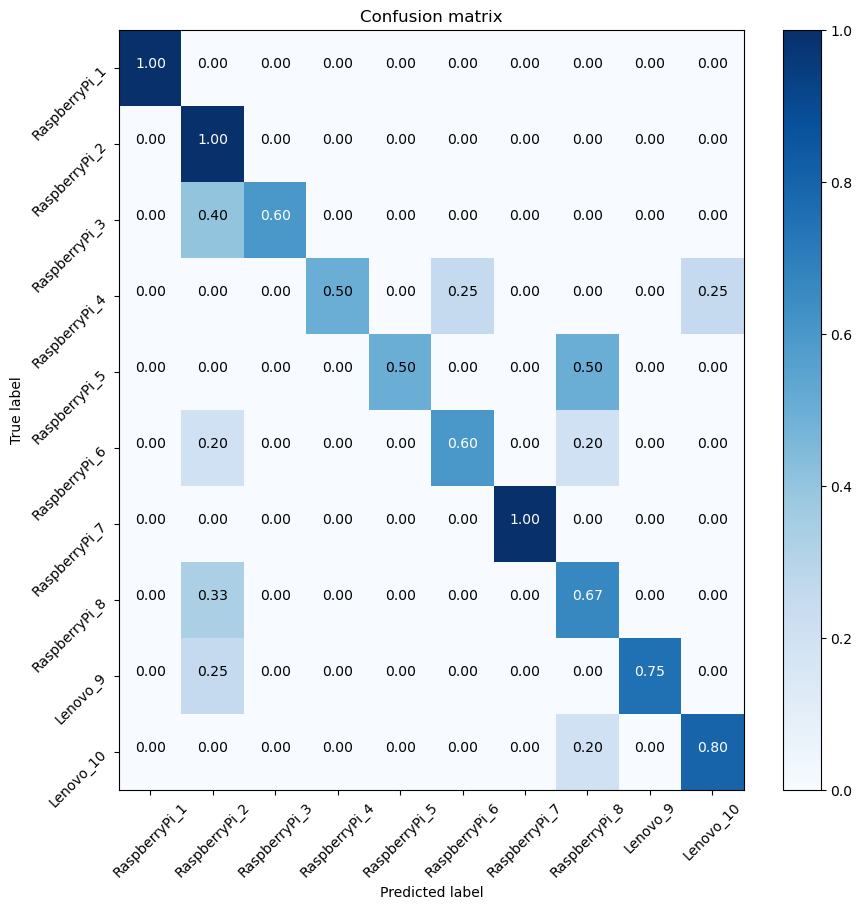}
         \caption{Oracle \cite{oracle} performance with the maximum supported sample length}
         \label{fig:oracle_ttsd}
     \end{subfigure}
     \begin{subfigure}[b]{0.32\textwidth}
         \centering
         \includegraphics[width=\textwidth, height=5 cm]{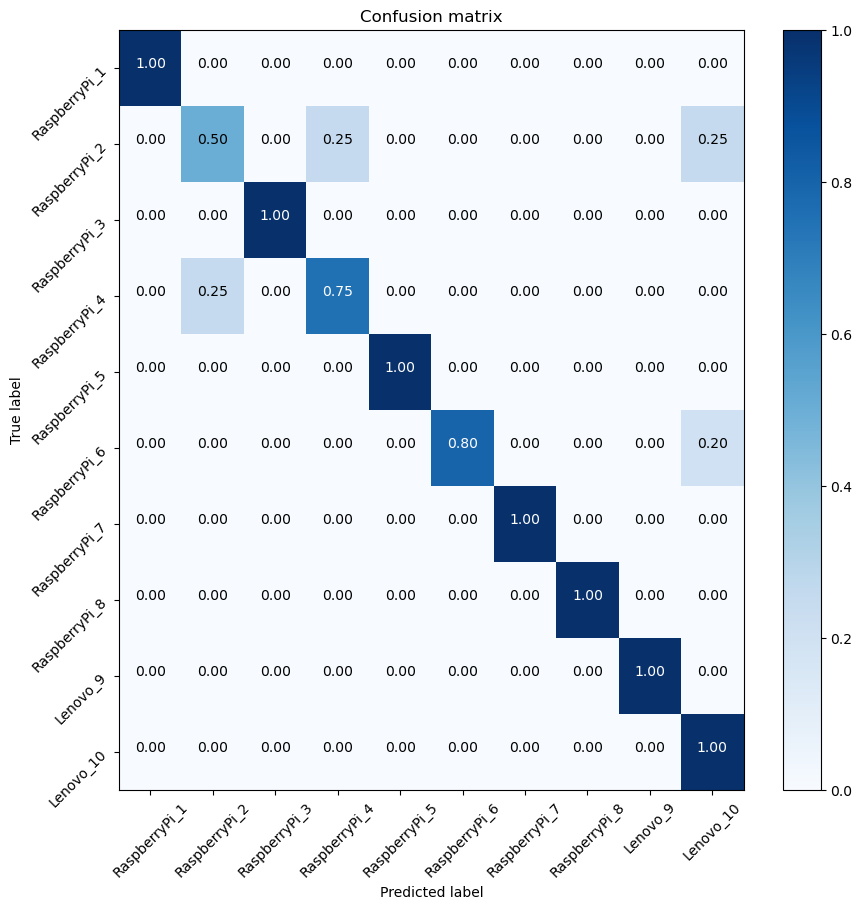}
         \caption{Mbed-ATN architecture performance with maximum supported sample length}
         \label{fig:attn_1M_ttsd}
     \end{subfigure}
     \hfill
        \caption{Achievable BT fingerprinting performance on TTS (same day same location testbed setup)}
        \label{fig:ttsd_cf}
\end{figure*}

\begin{figure*}[h!]
     \centering
      \begin{subfigure}[b]{0.32\textwidth}
         \centering
         \includegraphics[width=\textwidth, height=5 cm]{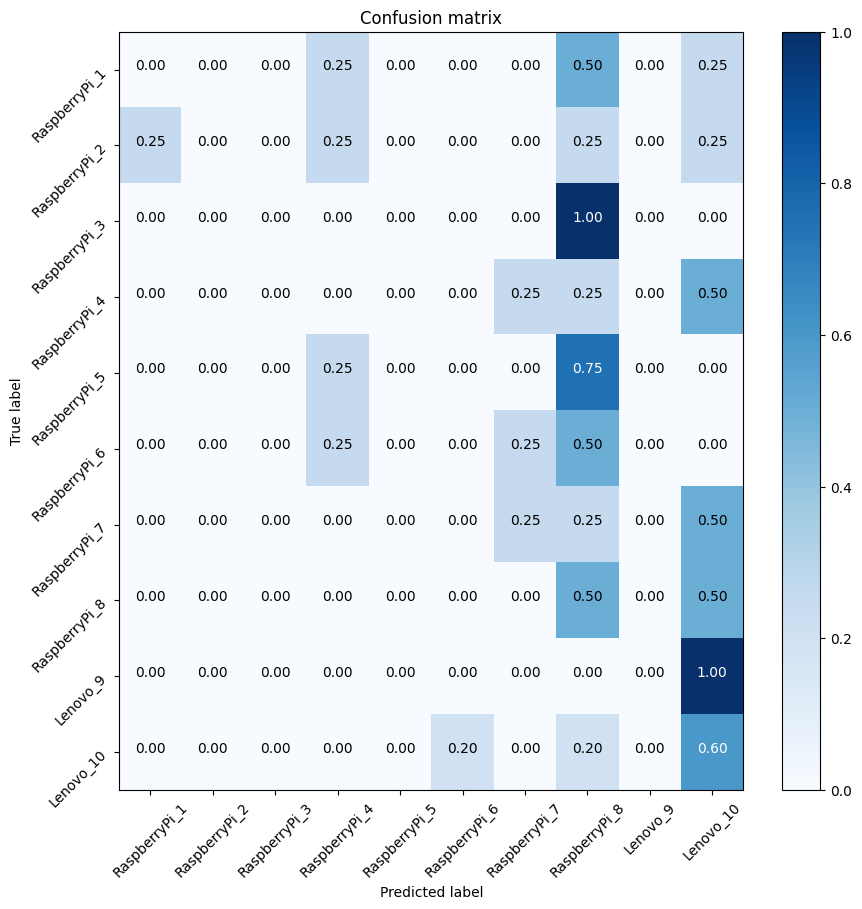}
         \caption{GRU network \cite{gru_rff_droy} performance with the maximum supported sample length}
         \label{fig:gru_ttddl}
     \end{subfigure}
     \hfill
     \begin{subfigure}[b]{0.32\textwidth}
         \centering
         \includegraphics[width=\textwidth, height=5 cm]{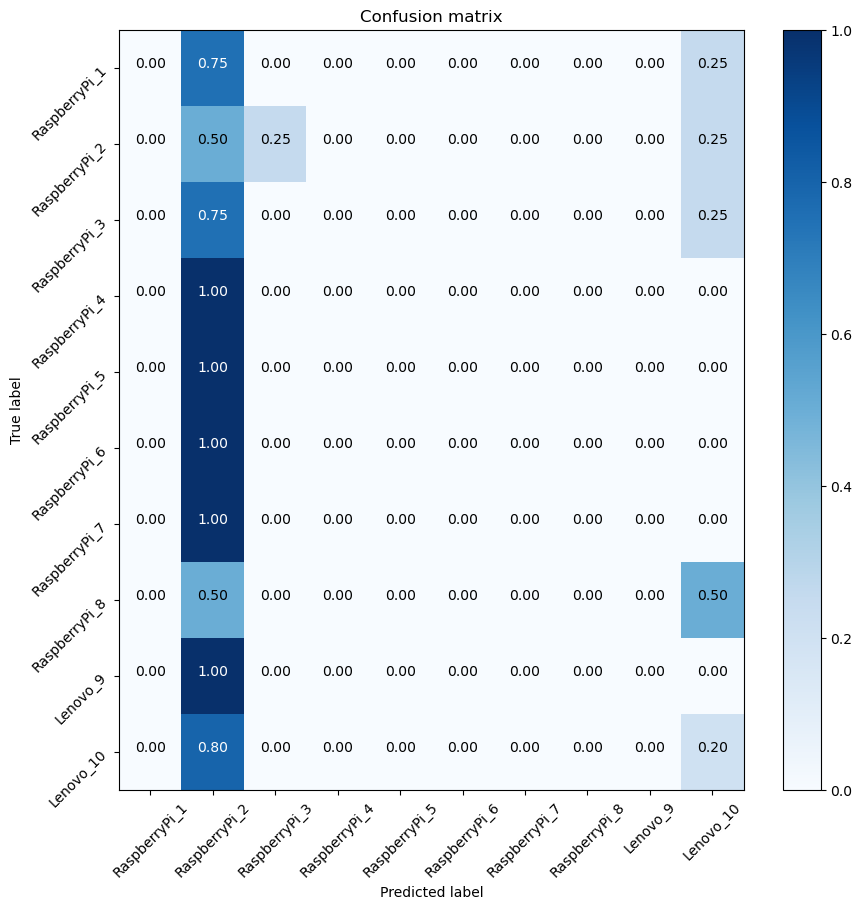}
         \caption{Oracle \cite{oracle} performance with the maximum supported sample length}
         \label{fig:oracle_ttddl}
     \end{subfigure}
     \hfill
     \begin{subfigure}[b]{0.32\textwidth}
         \centering
         \includegraphics[width=\textwidth, height=5 cm]{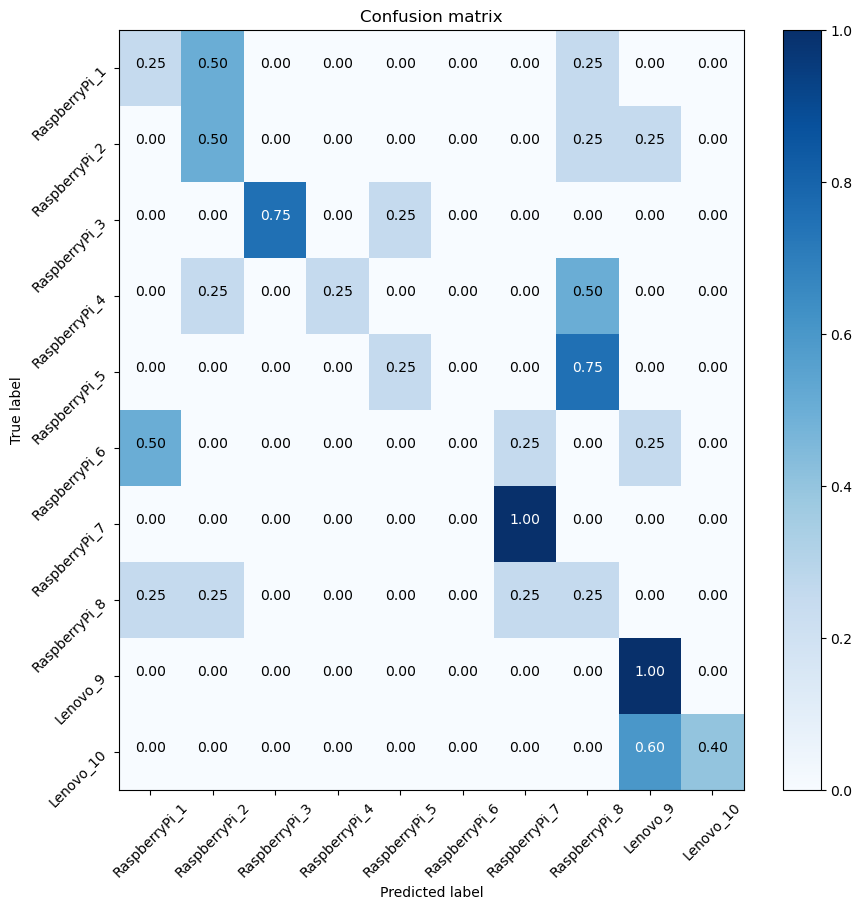}
         \caption{Mbed-ATN architecture performance with maximum supported sample length}
         \label{fig:attn_1M_ttddl}
     \end{subfigure}
     \hfill
        \caption{Achievable BT fingerprinting performance on TTD (different day different location testbed setup)}
        \label{fig:ttddl_cf}
\end{figure*}
\begin{figure}[h!]
     \centering
     \begin{subfigure}[b]{0.4\textwidth}
         \centering
         \includegraphics[width=\textwidth, height=4 cm]{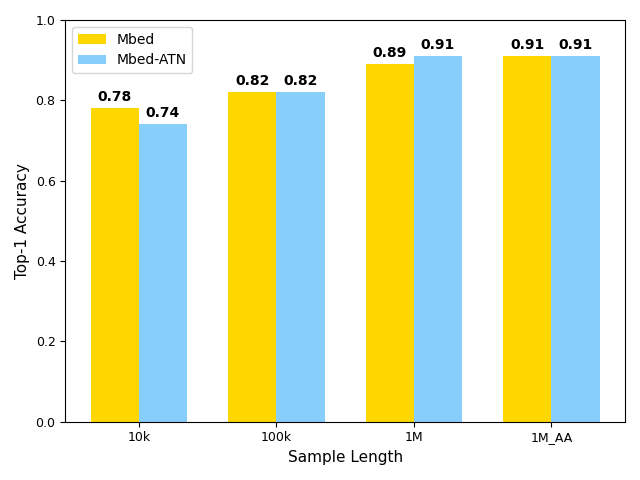}
         \caption{TTS Scenario}
         \label{fig:mbedVSattn_ttsd}
     \end{subfigure}
     \hfill
     \begin{subfigure}[b]{0.4\textwidth}
         \centering
         \includegraphics[width=\textwidth, height=4 cm]{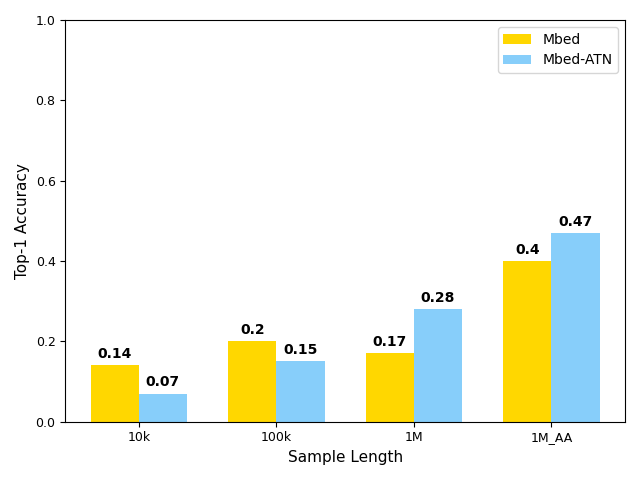}
         \caption{TTD Scenario}
         \label{fig:mbedVSattn_ttddl}
     \end{subfigure}
     \hfill
        \caption{Demonstrating effect of sample length and anti-aliasing decimation on Mbed-ATN fingerprinting performance.}
        \label{fig:mbedVSattn}
\end{figure}

In this study, we evaluate the effect of anti-aliasing decimation as opposed to straightforward downsampling. For this, we decimate the 40 MS capture to 1 MS with an anti-aliasing order 8 Chebyshev type I filter. Here, the waveform is subject to anti-aliasing filtering prior to downsampling. The effects of anti-aliasing decimation is shown in Table \ref{tab:BTanti}. Here, we can evidently see the performance increase of the models with reference to the downsampling without anti-aliasing filtering in Table \ref{tab:BT1M}. This implies that the plain decimation was causing aliasing thereby distorting the emitter signatures affecting their distinguishability. We show that Mbed-ATN achieves $23.1$\% higher TPR, $11.1$\% lesser false alarms, and $17.7$\% higher accuracy compared to the Mbed module under the TTD setting. To truly understand, the effect of anti-aliasing, we contrast the Mbed-ATN framework's performance under the TTD setting in Table \ref{tab:BT1M} and Table \ref{tab:BTanti}. Note the $99.5$\% increase in TPR, $25.6$\% drop in false alarms, and $69.1$\% spike in the top-1 accuracy of the Mbed-ATN with the anti-aliasing decimated samples. With respect to the TTD case in Table \ref{tab:BT10k}, the Mbed-ATN demonstrates a $5.32\times$ higher TPR, $37.9$\% fewer false positives, and $6.74\times$ higher accuracy with the increased sample length subject to anti-aliasing filtering. To shed more light on this visually, we depict this increase in accuracy in Fig.\ref{fig:mbedVSattn}. Here, the label 1M\_AA denotes a sample length of 1 MS with anti-aliasing decimation and we also show the accuracy with a sample length of $100$ kS. The improved generalization capability with longer sample length and anti-aliasing decimation under the challenging TTD setting with the adoption of the ATN module can be clearly seen in Fig.\ref{fig:mbedVSattn_ttddl}. This study shows the combined effect of higher sample length and anti-aliasing on the performance of the proposed Mbed-ATN fingerprinting framework.
Figures \ref{fig:ttsd_cf} and \ref{fig:ttddl_cf} show the confusion matrices of the benchmark models and proposed Mbed-ATN at their maximum supported input tensor lengths under the TTS and TTD experimental settings. The superiority of the proposed Mbed-ATN model in terms of the true positives, true negatives, false positives, and false negatives are evident in both the challenging TTD scenario and the TTS setup. These evaluations validate the improved generalization and lightweight nature of the proposed Mbed-ATN framework taking it one step closer to practical deployment. While this is a significant improvement in the TTD setting for BT compared to existing literature, we are currently working on other refinements to make the TTD metrics close to the TTS. Furthermore, we also share the dataset so that the larger research community can continue to work on improving the generalization capability of RF fingerprint approaches using real-world data \cite{AJagannath2022_BT_port}.   

\subsection{Why not raw IQ samples ?}
\begin{figure}[htp]
\centering
\includegraphics[width=0.99\columnwidth, height=5 cm]{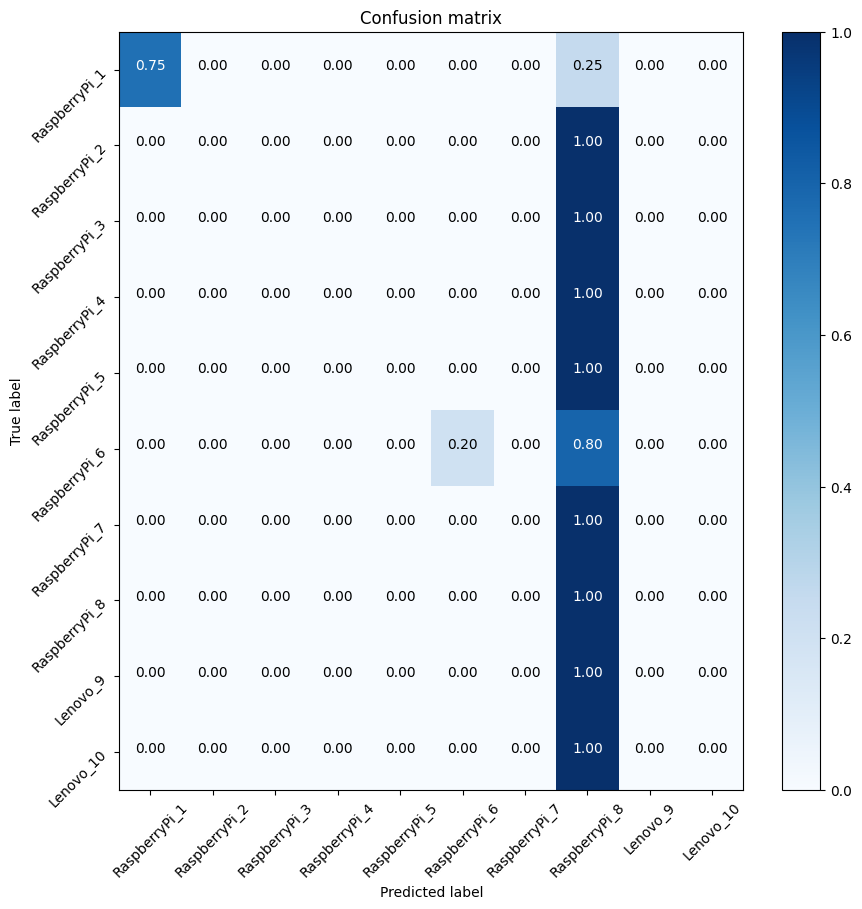} 
\caption{Performance using raw IQ samples under a TTS setup.}
\label{fig:iq_cm}
\end{figure}

We also validate the necessity to use the extracted features as in the expression \ref{eq:ip_format} in section \ref{sec:mbed_atn} as input rather than raw IQ samples. To achieve this, we used raw IQ samples arranged as a tensor of length 1 MS with two rows containing I and Q samples separately in each row as input to the Mbed-ATN architecture under the TTS setup. The confusion matrix of the architecture with raw IQ samples is shown in Fig.\ref{fig:iq_cm}. The false predictions and the inability of the model to distinguish between the emitters can be clearly seen from the confusion matrix. The need for the appropriate input tensor is obvious in comparison to the confusion matrix in Fig.\ref{fig:attn_1M_ttsd} which has only negligible false predictions. We also tabulate the performance metrics and provide a direct comparison with the Mbed-ATN using the appropriate input tensor as in expression \ref{eq:ip_format} in Table \ref{tab:BTip}. Adopting the appropriate input format resulted in a top-1 accuracy of 91\% and FPR of 1.1\%.
\begin{table}
\centering
\caption{Fingerprinting performance under various input formats (sample length = 1M). \label{tab:BTip}}
\begin{tcolorbox}[width=7.1 cm,tab2,tabularx={p{2.6 cm}|p{0.8 cm}|p{0.8 cm}|p{0.8 cm}}]
\textbf{Model} & \textbf{TPR} &\textbf{FPR} & \multicolumn{1}{c}{\textbf{Top-1 Acc.}} \\
\hline
Mbed-ATN (IQ) &0.166 &0.089 &0.195\\
Mbed-ATN (Tensor \ref{eq:ip_format}) &\textbf{0.905} &\textbf{0.011} &\textbf{0.905}\\\hline
\end{tcolorbox}
\vspace{-.2 cm}
\end{table}
\section{Conclusion}

We proposed and presented a detailed analysis of Mbed-ATN, an embedding-assisted attentional framework for enhancing the generalization capability of the fingerprinting architecture. The proposed model is scalable for supporting large input tensor lengths of $1$ MS while using significantly less GPU memory. The proposed Mbed-ATN utilizes $65.2\times$ lesser memory in contrast to the state-of-the-art Oracle architecture for an input length of $M=100$ kS. Further, for a 10 kS sample length, the Mbed-ATN utilizes $16.9\times$ fewer FLOPs and $7.5\times$ fewer trainable parameters with respect to Oracle. We showed that the inference time of the proposed Mbed-ATN is $21.21\times$ lesser than that of the benchmark GRU model and attains a $1.15\times$ and $9.17\times$ lesser memory usage at sample lengths of $M=10$ kS and $M=100$ kS respectively when compared to GRU network. A detailed empirical study on the effect of higher sample length and anti-aliasing decimation was demonstrated for the proposed Mbed-ATN framework in showcasing the improved generalization capability of the model with the introduction of attentional learning. Unlike the existing literature, we resorted to the challenging different time frame, location, and experimental setup (TTD) scenarios along with demonstrating the GPU efficiency of the model in validating the real-world deployment merit of the proposed framework.
\bibliography{bibfile.bib}
\bibliographystyle{IEEEtran}

\begin{IEEEbiography}[{\includegraphics[width=1in,height=1.25in,clip,keepaspectratio]{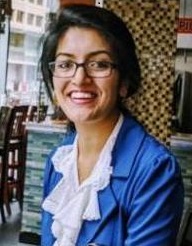}}]{Anu Jagannath} is the Founding Associate Director of Marconi-Rosenblatt AI/ML Innovation Lab at ANDRO Computational Solutions, LLC. She received her MS degree from State University of New York at Buffalo in Electrical Engineering. She is also a part-time PhD candidate and is with the Institute for  the Wireless Internet of Things at Northeastern University, USA. Her research focuses on deep earning, reinforcement learning, adaptive signal processing, software defined radios, and signal intelligence. She has rendered her reviewing service for several leading IEEE conferences and Journals. She is an IEEE Senior Member. She is the co-Principal Investigator (co-PI) and Technical Lead in multiple Rapid Innovation Fund (RIF) and SBIR/STTR efforts involving  applied AI/ML for wireless communications. She is also the co-inventor of 11 US Patents (granted and pending).
\end{IEEEbiography}

\begin{IEEEbiography}[{\includegraphics[width=1in,height=1.25in,clip,keepaspectratio]{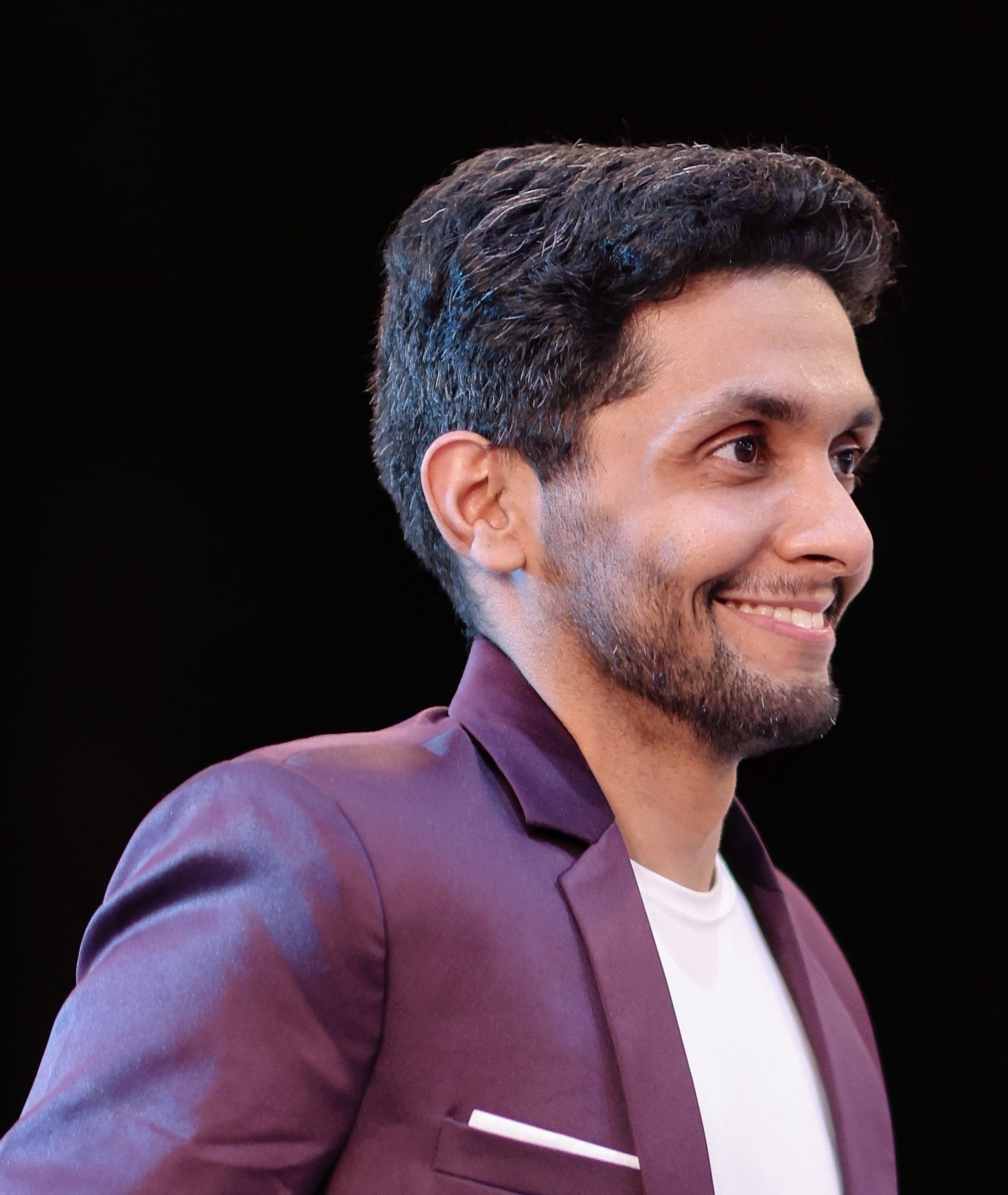}}]{Jithin Jagannath} is the Chief Scientist of Technology and Founding Director of the Marconi-Rosenblatt AI/ML Innovation Lab at ANDRO Computational Solutions. He is also the Adjunct Assistant Professor in the Department of Electrical Engineering at the University at Buffalo. He received his Ph.D. degree in Electrical Engineering from Northeastern University. He has been the Principal Investigator in research efforts for several customers including the U.S. Army, U.S. Navy, U.S. SOCOM, DHS, and AFSOR in the field of Beyond 5G, signal processing, RF signal intelligence, cognitive radio, cross-layer ad-hoc networks, Internet-of-Things, AI-enabled wireless, and machine learning. 

Dr. Jagannath is an IEEE Senior member and has been invited as Keynote Speaker and Panelist discussing his vision at leading IEEE Conferences. He serves as a full member of the IEEE SPS Applied Signal Processing Systems Technical Committee and several other Technical Program Committees (TPCs). Additionally, he serves on the editorial board of Computer Networks (Elsevier). Dr. Jagannath's recent research has led to over 40 peer-reviewed journal and conference publications. He is the co-inventor of 16 U.S. Patents (granted and pending). Dr. Jagannath was the recipient of 2021 IEEE Region 1 Technological Innovation Award with the citation. He is also the recipient of AFCEA International Meritorious Rising Star Award for achievement in engineering and AFCEA 40 Under 40 award. 
\end{IEEEbiography}

\vfill
\end{document}